\documentclass[prb,amsmath,twocolumn,amssymb,superscriptaddress,showpacs]{revtex4} 
\usepackage{bm}
\usepackage[]{graphicx}
\usepackage[tight]{units}
\usepackage{color}
\usepackage{ulem}
\usepackage[colorlinks=true,citecolor=blue,linkcolor= blue,linkcolor=blue,urlcolor=blue]{hyperref}

\newcommand{\cip}{$\sigma_{_{\|}}$~}
\newcommand{\cpp}{$\sigma_{_{\perp}}$~}
\newcommand{\PFip}{$\text{PF}_{_{\|}}$ ~}
\newcommand{\PFpp}{$\text{PF}_{_{\perp}}$~}

\newcommand{\PFipx}{$\text{PF}_{_{\|}}$}
\newcommand{\PFppx}{$\text{PF}_{_{\perp}}$}

\newcommand{\Sip}{$S_{_{\|}}$~}
\newcommand{\Spp}{$S_{_{\perp}}$~}
\newcommand{\ratio}{$\nicefrac{\sigma_{\|}}{\sigma_{\perp}}$}
\newcommand{\Sipx}{S_{_{\|}}}
\newcommand{\Sppx}{S_{_{\perp}}}
\newcommand{\BiTe}{$\text{Bi}_2\text{Te}_3$~}
\newcommand{\SbTe}{$\text{Sb}_2\text{Te}_3$~}
\newcommand{\SBSL}{$\text{Bi}_2\text{Te}_3/\text{Sb}_2\text{Te}_3$-SL~}

\newcommand{\BiTex}{$\text{Bi}_2\text{Te}_3$}
\newcommand{\SbTex}{$\text{Sb}_2\text{Te}_3$}
\newcommand{\SBSLx}{$\text{Bi}_2\text{Te}_3/\text{Sb}_2\text{Te}_3$-SL}

\newcommand{\f}[1]{fig.~\ref{fig:#1}}
\newcommand{\F}[1]{Figure \ref{fig:#1}}
\newcommand{\Ff}[1]{figure \ref{fig:#1}}

\newcommand{\V}{Venkatasubramanian \textit{et al.} ~}

\begin{document}

\title[]{Influence of strain on anisotropic thermoelectric transport of \BiTe and \SbTe}
\author{N. F. Hinsche}
\email{nicki.hinsche@physik.uni-halle.de}
\affiliation{Institut f\"{u}r Physik, Martin-Luther-Universit\"{a}t Halle-Wittenberg, D-06099 Halle, Germany}
\author{B. Yu. Yavorsky}
\affiliation{Institut f\"{u}r Physik, Martin-Luther-Universit\"{a}t Halle-Wittenberg, D-06099 Halle, Germany}
\author{I. Mertig}
\affiliation{Institut f\"{u}r Physik, Martin-Luther-Universit\"{a}t Halle-Wittenberg, D-06099 Halle, Germany}
\affiliation{Max-Planck-Institut f\"{u}r Mikrostrukturphysik, Weinberg 2, D-06120 Halle, Germany}
\author{P. Zahn}
\affiliation{Institut f\"{u}r Physik, Martin-Luther-Universit\"{a}t Halle-Wittenberg, D-06099 Halle, Germany}
\date{\today}

\begin{abstract}
 On the basis of detailed first-principles calculations and semi-classical Boltzmann transport, 
 the anisotropic thermoelectric transport properties of \BiTe and \SbTe under strain were investigated. 
 It was found that due to compensation effects of the strain dependent thermopower and electrical 
 conductivity, the related powerfactor will decrease under applied in-plane strain for \BiTex, while being stable for 
 \SbTex. 
 A clear preference for thermoelectric transport under hole-doping, as well as for the in-plane transport 
 direction was found for both tellurides. In contrast to the electrical conductivity anisotropy, the 
 anisotropy of the thermopower was almost robust under applied strain. 
 The assumption of an anisotropic relaxation time for \BiTe suggests, 
 that already in the single crystalline system strong anisotropic scattering effects should play a role.
\end{abstract}

\pacs{31.15.A-,71.15.Mb,72.20.Pa,72.20.-i}

\maketitle


\section{Introduction}
Thermoelectric (TE) materials are used as solid state energy devices which convert waste heat into electricity or 
electrical power directly into cooling or heating \cite{Sales:2002p6580,Majumdar:2004p6568,Bottner:2006p2812}. 
Telluride based thermoelectrics, e.g. the bulk materials bismuth (\BiTex) and antimony telluride (\SbTex) and their related alloys, dominate efficient 
TE energy conversion at room temperature 
for the last 60 years \cite{Goldsmid:1958p15284,Venkatasubramanian:2001p114}. The materials TE efficiency is quantified 
by the figure of merit
\begin{equation}
ZT=\frac{\sigma S^{2}}{\kappa_{el} + \kappa_{ph}} T,
\label{eq1}
\end{equation}
where $\sigma$ is the electrical conductivity, $S$ the thermopower, $\kappa_{el}$  and 
$\kappa_{ph}$ are the electronic and phononic contribution to the thermal conductivity, respectively. 
From Eq.~\ref{eq1} it is obvious, that a higher ZT is obtained by decreasing the denominator 
or by increasing the numerator, the latter being called powerfactor $\text{PF}=\sigma S^{2}$. 
While bulk \BiTe and \SbTe show $Z$T values smaller 1 and applications have been
limited to niche areas, a break-trough experiment of \V
showed a remarkable $ZT = 2.4/1.5$ for p-type/n-type superlattices (SL) composed of the two bulk tellurides \cite{Venkatasubramanian:1999p13956,Venkatasubramanian:2000p7305,Venkatasubramanian:2001p114}.
With the availability of high-ZT materials, many new applications will emerge \cite{Majumdar:2004p6568}.
The idea of thermoelectric SL follows the idea of phonon-blocking and electron-transmitting at the same time.
It suggests that cross-plane transport along the direction 
perpendicular to the artificial interfaces 
of the SL reduces phonon heat conduction while maintaining or even enhancing the electron transport \cite{Bottner:2006p2812}. 
While some effort in experimental research was done \cite{Beyer:2002p15267,Bottner:2004p6591,Konig:2011p48,Liao:2010p11008,Peranio:2006p15247,Touzelbaev:2011p15270}, only a few theoretical works discuss the possible transport 
across such SL structures\cite{Park:2010p11006,Li:2004p15238}. While Park \textit{et al.}\cite{Park:2010p11006} discussed the effect of 
volume change on the in-plane thermoelectric transport properties of \BiTex, \SbTe and their related compound,  Li \textit{et al.}\cite{Li:2004p15238} 
focussed on the calculation of the electronic structure for a \SBSLx, stating changes of the mobility anisotropy estimated 
from effective masses. 

Superlattices are anisotropic by definition 
and even the telluride bulk materials show intrinsic anisotropic structural and electronic properties. 
However, investigations of \V found a strong decrease 
for the mobility anisotropy and the thermoelectric properties for the \SBSL at certain periods. 
The reason for this behaviour is still on debate and could be related to 
strain effects which are induced by the epitaxial growth of the \SBSLx.
To extend previous works~\cite{Scheidemantel:2003p14961,Thonhauser:2003p14996,Huang:2008p559} 
and to clarify the open question on the reduced anisotropy, we are going to discuss in this paper 
the anisotropic electronic transport in bulk \BiTe and \SbTe and the possible influence of strain in epitaxially 
grown SL on the TE properties. 

\vspace{0.3cm}
For this purpose the paper will be organized as follows. 
In section \ref{method} we introduce our first principle 
electronic structure calculations based on density functional theory and the semi-classical transport calculations 
based on the solution of the linearized Boltzmann equation. With this, we discuss the thermoelectric 
transport properties, that is electrical conductivity, thermopower and the related powerfactor, of unstrained \BiTe and \SbTe 
with a focus on their directional anisotropies. While in epitaxially grown \SBSL
the atoms near the interfaces may be 
shifted from their bulk positions due to the lattice mismatch and the changed local environment, 
we modelled \BiTe with the
experimental lattice parameters and interatomic distances of \SbTe, and
vice versa. We assume that from these two limiting cases one could estimate
the effect of the interface relaxation on the electronic and transport
properties in \SBSLx. 
With that structural data we first analyse in section \ref{unstrained} the anisotropic thermoelectric properties of the unstrained bulk 
systems, while in section \ref{strained} a detailed view on the influence of strain, which may occur in \SBSLx, 
on the electronic transport of these tellurides is given. Throughout the paper we quote \BiTe (\SbTex) as strained, 
if it is considered in the lattice structure of \SbTe (\BiTex). 
As in the SL p-type, as well as n-type, transport was reported, we discuss the concentration dependence for both types of 
carriers on the transport properties.

\section{\label{method} Methodology}

For both bismuth and antimony telluride we used the experimental lattice
parameters and relaxed atomic positions \cite{Landolt} as provided for the rhombohedral crystal structure
with five atoms, i.e. one formula unit, per unit cell belonging to the space
group $D^5_{3d}$ ($R\bar{3}m$). The related layered hexagonal structure is composed out of three 
formula units and has the lattice parameters 
${a^{hex}_{BiTe}}=4.384${\AA ,} $c^{hex}_{BiTe}=30.487${\AA ,} and
${a^{hex}_{SbTe}}=4.264${\AA ,} $c^{hex}_{SbTe}=30.458${\AA ,} for
\BiTe and \SbTex, respectively. In fact, the main difference between the lattices of 
\BiTe and \SbTe is a decrease of the in-plane lattice constant with an accompanied 
decrease in cell volume. So, a change between the two lattice constants can be related to either compressive or tensile 
in-plane strain. This is very similiar to the approach by Park \textit{et al.}\cite{Park:2010p11006}, while omitting 
computational relaxation of internal atomic positions. 

Our electronic structure calculations are performed in two steps. In a first step the detailed band structure of the strained and unstrained \BiTe 
ans \SbTe was obtained by first principles density functional theory calculations (DFT), 
as implemented in the fully relativistic screened Korringa-Kohn-Rostoker Greens-function method (KKR) \cite{Gradhand:2009p7460}. Within this approach the 
\textsc{Dirac}-equation is solved self-consistently and with that spin-orbit-coupling is included. 
Exchange and correlation effects were accounted for by the local density approximation (LDA) parametrized by Vosco, Wilk, and
Nusair \cite{Vosko1980}. A detailed discussion on the influence of strain on the band structure topology of \BiTe and 
\SbTe is recently published \cite{BY}. 
\begin{figure}[t]
\centering
\includegraphics[width=0.49\textwidth]{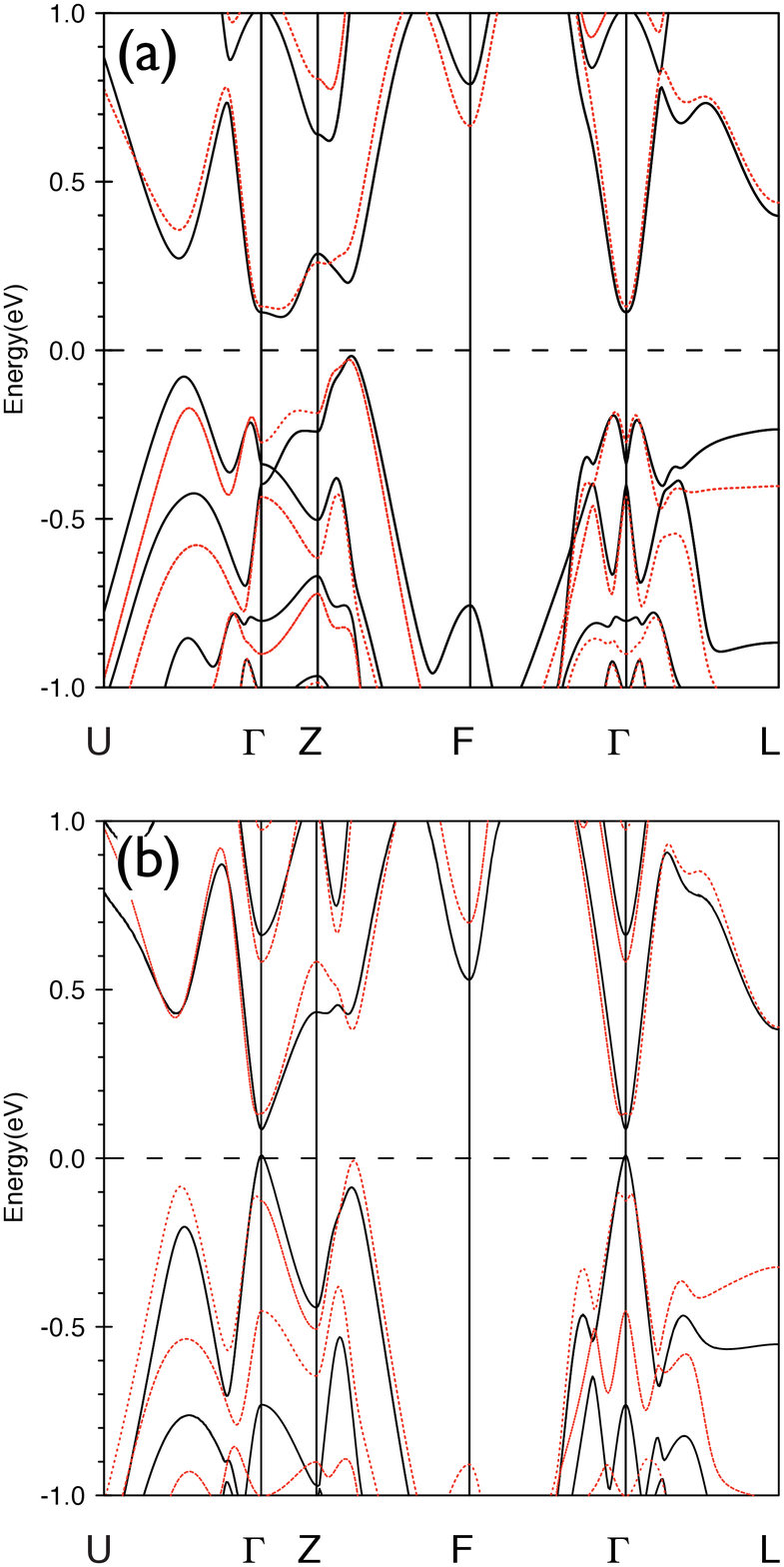}
\caption{\label{fig:0}(color online) Band structures of (a) Bi$_2$Te$_3$ and (b) Sb$_2$Te$_3$
along symmetry lines for both unstrained (black solid lines) and strained (red dashed lines) 
lattices. Energies are given relative to the valence band maximum.}
\end{figure}
With the well converged results from the first step we obtain 
the thermoelectric transport properties 
by solving the linearized Boltzmann equation in relaxation time approximation (RTA) within an in-house developed Boltzmann 
transport code \cite{Mertig:1999p12776,Zahn:1995p14971,Hinsche:2011p15276}. 
Boltzmann transport calculations for thermoelectrics have been carried out for quite a long time and show 
reliable results for metals~\cite{Vojta:1992p1395,Yang:2008p11828,Barth:2010p15131} as well as for wide- and narrow gap semiconductors~\cite{Singh:2010p14285,Parker:2010p13171,May:2009p14962,Lee:2011p14982,Hinsche:2011p15276}. TE transport calculations 
for bulk \BiTe \cite{Huang:2008p559,Lee:2006p1608,Park:2010p11006,Situmorang:1986p15020} and \SbTe \cite{Thonhauser:2004p15235,Thonhauser:2003p14996,Park:2010p11006} were presented before.
Here the relaxation time $\tau$ is assumed to be constant with respect to wave vector k and energy on the scale of $k_{B}T$. 
This assumption is widely accepted for metals and highly doped semiconductors. Most of the presented results are in this high-doping regime. 
Within the RTA the transport distribution function  $\mathcal{L}_{\perp, \|}^{(0)}(\mu, 0)$ (TDF) ~\cite{Mahan:1996p508} and with this the 
generalized conductance moments $\mathcal{L}_{\perp, \|}^{(n)}(\mu, T)$ are defined as 
\begin{eqnarray}
& \mathcal{L}_{\perp, \|}^{(n)}(\mu, T)= \nonumber \\
&\frac{\tau_{\|,\perp}}{(2\pi)^3} \sum \limits_{\nu} \int\ d^3k \left( v^{\nu}_{k,(\perp, \|)}\right)^2 (E^{\nu}_k-\mu)^{n}\left( -\frac{\partial f_{(\mu,T)}}{\partial E} \right)_{E=E^{\nu}_k} \nonumber .
\\
\label{Tcoeff}
\end{eqnarray} 
$v^{\nu}_{k,(\|)}$, $v^{\nu}_{k,(\perp)}$ denote the group velocities in the directions in the hexagonal basal plane and perpendicular  to it, respectively. 
Within here the group velocities were obtained as derivatives along 
the lines of the Bl\"ochl mesh in the whole Brillouin zone\cite{BY}. 
A detailed discussion on implications and difficulties on the numerical determination 
of the group velocities in highly anisotropic materials is currently published elsewhere\cite{PZ}. 
As can be seen straight forwardly, the electrical conductivity $\sigma$ in- and cross-plane is then given by
\begin{eqnarray}
\sigma_{_{\perp, \|}}=2e^2 \mathcal{L}_{\perp, \|}^{(0)}(\mu, T)
\label{Sigma}
\end{eqnarray}
and the temperature- and doping-dependent thermopower states as
\begin{eqnarray}
S_{_{\perp, \|}}=\frac{1} {eT} \frac{\mathcal{L}_{\perp, \|}^{(1)}(\mu,T)} {\mathcal{L}_{\perp, \|}^{(0)}(\mu,T)}
\label{Seeb}
\end{eqnarray}
for given 
chemical potential $\mu$ at temperature $T$ and extrinsic carrier concentration $N$ determined by an integration 
over the density of states $n(E)$
\begin{eqnarray}
N=\int \limits_{\mu-\Delta E}^{\text{VB}^{max}} \text{d}E \,  n(E) [f_{(\mu,T)}-1]+
\int \limits_{\text{CB}^{min}}^{\mu+\Delta E} \text{d}E \, n(E) f_{(\mu,T)}
\label{Dop},
\end{eqnarray}
where $\text{CB}^{min}$ is the conduction band minimum and $\text{VB}^{max}$ is the 
valence band maximum. The energy range $\Delta E$ has to be taken sufficiently large to cover the tails 
of the \textsc{Fermi-Dirac} distribution function $f_{(\mu,T)}$ and to ensure convergence of the integrals in eq.~\ref{Tcoeff} and \ref{Dop} \cite{Hinsche:2011p15276}.
The k-space integration of eq.~\ref{Tcoeff} for a system with an intrinsic anisotropic texture is quite demanding. 
In previous publications \cite{PZ,BY} we stated on the relevance of adaptive integration methods needed to reach 
convergence of the energy dependent TDF. Especially in regions close to the band edges the anisotropy of 
the TDF requires a high density of the k-mesh. Here, convergence tests 
for the transport properties showed that at least 150 000 k-points in the 
entire BZ had to be included for sufficient high doping rates ($N \geq \unit[1\times 10^{19}]{cm^{-3}}$), while for energies 
near the band edges even more than 56 million k-points were required to reach the 
analytical values for the conductivity anisotropies at the band edges\footnote{The analytical value of the ratio \nicefrac{\cip}{\cpp} at the band edges 
was obtained by scanning the 
energy landscape near the conduction band minimum and valence band maximum fitting the dispersion relation in terms of an effective mass tensor. 
A detailed description is given in a recent publication by Ref.~\onlinecite{BY}}.
Within the RTA, from comparison of the calculated electrical conductivities (eq.~\ref{Sigma}) with experiment it is possible to conclude 
on the directional anisotropy of $\tau$. For the thermopower S (eq.~\ref{Seeb}) the dependence of the TDF on the energy is essential. 
That is, not only the sloop of the TDF, moreover the overall functional behaviour of the TDF on the considered energy scale has to change to observe an impact 
on the thermopower. 
The calculations in this paper aim to cover band structure effects and not scattering specific impacts by an energy- and state-dependent relaxation time.
%

\section{\label{unstrained}Anisotropic thermoelectric properties of unstrained \BiTe and \SbTe}

In order to understand the experimental findings on the in-plane and cross-plane transport of the \SBSLx, in the 
following section the anisotropies 
of the electrical conductivity, the thermopower 
and the related powerfactor of bulk \BiTe and \SbTe are discussed. 
Even though the behaviour of \SbTe is strongly p-type with a extrinsic carrier concentration of $N = \unit[1 \dots 10\times 10 ^{20}]{cm^{-3}}$,\cite{MRowe:1995p15325} 
we also discuss the related n-doped case, as in \SBSL n- as well as p-doping was reported. Bulk \BiTe is 
known to be inherent electron conducting, while hole doping is experimentally achievable for bulk systems \cite{Delves:1961p7491,Goldsmid:1958p15284,Jeon:1991p13910,Kaibe:1989p15019}.
\begin{figure}[t]
\centering
\includegraphics[width=0.49\textwidth]{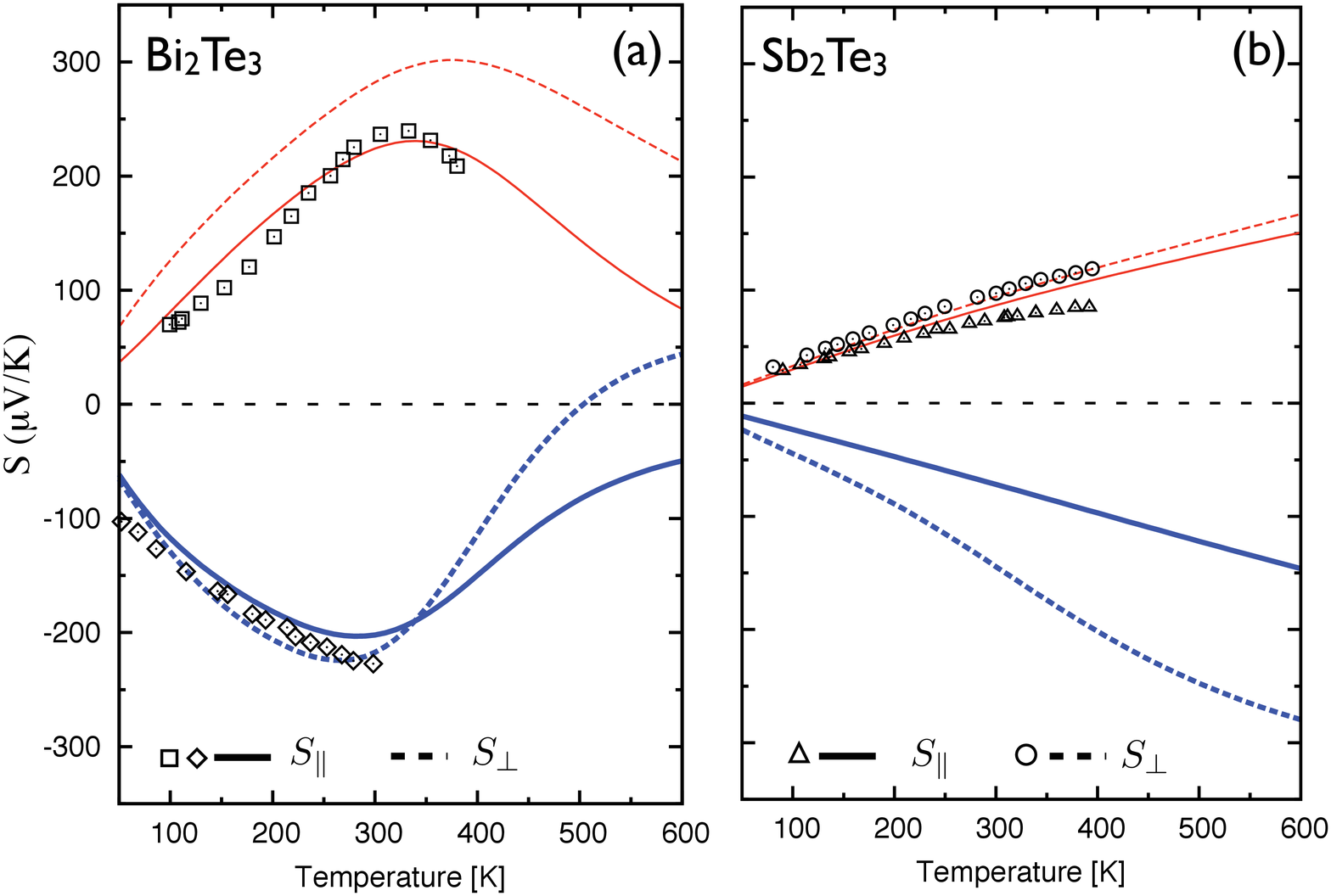}
\caption{\label{fig:1}(color online) Anisotropic thermopower for bulk (a) \BiTe and (b) \SbTe in their unstrained bulk lattice constants. Electron doping refers to the blue (thick) lines
 in the lower part of the figure, while red (thin) lines refer to hole 
doping and positive values of the thermopower. Solid lines show the in-plane part \Sip of the thermopower, while dashed lines show the cross-plane part \Spp. 
The extrinsic charge carrier concentration of \BiTe and \SbTe was fixed to $N = \unit[1\times 10^{19}]{cm^{-3}}$ 
and $N = \unit[1\times 10 ^{20}]{cm^{-3}}$, respectively. 
Experimental data (squares, diamonds, circles, triangles) from Ref.~\onlinecite{Stordeur:1975p15016,Kaibe:1989p15019,Stordeur:1976p15137} 
are given for comparison.}
\end{figure}
\F{1} shows the variation of the anisotropic thermopower for unstrained \BiTe and \SbTe in a wide temperature range. 
The extrinsic charge carrier concentration of \BiTe and \SbTe was fixed to $N = \unit[1\times 10^{19}]{cm^{-3}}$ 
and $N = \unit[1\times 10 ^{20}]{cm^{-3}}$, respectively. As a reference experimental values for both single crystalline materials at 
the same doping conditions are given and an excellent agreement can be stated. It is worth noting, that within eq.~\ref{Seeb} the calculation 
of the thermopower is completely free of parameters. For \BiTe the in-plane thermopower reaches a maximum of $\Sipx \sim \unit[-200]{\mu V/K}$  
at $\unit[300]{K}$, while the maximum for the hole-doped case is shifted to slightly higher temperatures of $\unit[350]{K}$ with a maximum values of 
$\Sipx \sim \unit[225]{\mu V/K}$. We note, that the temperature of the maximum is slightly overestimated. This might be caused by the missing temperature 
dependence of the energy gap, which was determined as $E_g=\unit[105]{meV}$ for unstrained \BiTex. The anisotropy of the thermopower is more pronounced for 
the p-doped case. Here the cross-plane thermopower \Spp is for the given doping always larger than the in-plane part \Sip. 
The anisotropy $\nicefrac{\Sipx}{\Sppx}$  is about $0.64$ at 100K, evolving to $\nicefrac{\Sipx}{\Sppx} \sim 0.79$ 
and $\nicefrac{\Sipx}{\Sppx} \sim 0.55$ at 300K and 500K, respectively. The sole available experimental data show no noticeable anisotropy for the 
thermopower in the hole-doped case \cite{Stordeur:1975p15016}.
For the electron-doped case the situation is more sophisticated. While upto 340K the overall anisotropy is rather small, with values $\nicefrac{\Sipx}{\Sppx} \sim 0.9$, 
a considerable decrease of \Spp at higher temperatures leads to high values of $\nicefrac{\Sipx}{\Sppx}$ for temperatures above 400K. This tendency could also be 
revealed by experiments \cite{zhit,Mueller1998}. 
The crossing point of \Sip and \Spp near room temperature could explain the fact of varying measured anisotropies for the thermopower at 300K. 
Here anisotropy ratios of  $\nicefrac{\Sipx}{\Sppx} = 0.97 \dots 1.10$ were reported \cite{Kaibe:1989p15019,Mueller1998}. The maximum peak of the thermopower near room temperature 
can be explained by the position of the chemical potential $\mu$ as a function of temperature at a fixed carrier concentration. 
For T much smaller than 300K 
the chemical potential is located in either the conduction- or valence band with the tails of the Fermi-Dirac-distribution in eq.~\ref{Tcoeff} only 
playing a subsidiary role. For rising temperatures the chemical potential shifts towards the band edges and $S$ maximizes. At these conditions the conduction is mainly unipolar. 
For higher temperatures the chemical potential shifts into the bandgap and conduction becomes bipolar leading to a reduced thermopower.
For the case of \SbTe, shown in \f{1}(b), the situation is different. Due to the ten times higher inherent doping and the smaller energy gap of $E_g=\unit[90]{meV}$, the chemical potential is located deeply
in the bands for the whole relevant temperature range. Therefore the functional behaviour can be understood in terms of the well known \textsc{Mott} relation, where 
equation \ref{Seeb} qualitatively coincides with $S \propto T \cdot \frac{\text{d} \ln \sigma(E)}{\text{d}E}|_{E=\mu} $ for the thermopower in RTA \cite{Cutler:1969p3655}. 
With increasing temperature the thermopower increases almost linearly, showing values of $\Sipx \sim \unit[87]{\mu V/K}$ and $\Sipx \sim \unit[-72]{\mu V/K}$ at 
300K for p- and n-doping, respectively. The anisotropy of the thermopower for the hole-doped case is around 
$\nicefrac{\Sipx}{\Sppx} = 0.91 $, almost temperature-independent and slightly underestimates the available 
experimental values \cite{Simon:1981p15250,Langhammer:1982p15249}. 
While for the electron-doped case the absolute values of the in-plane 
thermopower are comparable to those of the hole-doped case, the anisotropies are rather large. The anisotropy varies only weakly on temperature showing 
$\nicefrac{\Sipx}{\Sppx} = 0.48 \dots 0.52 $ over the hole temperature range. While bulk \SbTe states a strong p-character due to inherent defects, we note here again, that n-doping is available in heterostructures combining \BiTe and \SbTe \cite{Venkatasubramanian:2001p114}. 

A strongly enhanced cross-plane thermopower \Spp could lead to a strongly enhanced 
powerfactor \PFppx, if the cross-plane electrical conductivity \cpp is maintained at the bulk value.
\begin{figure}[t]
\centering
\includegraphics[width=0.49\textwidth]{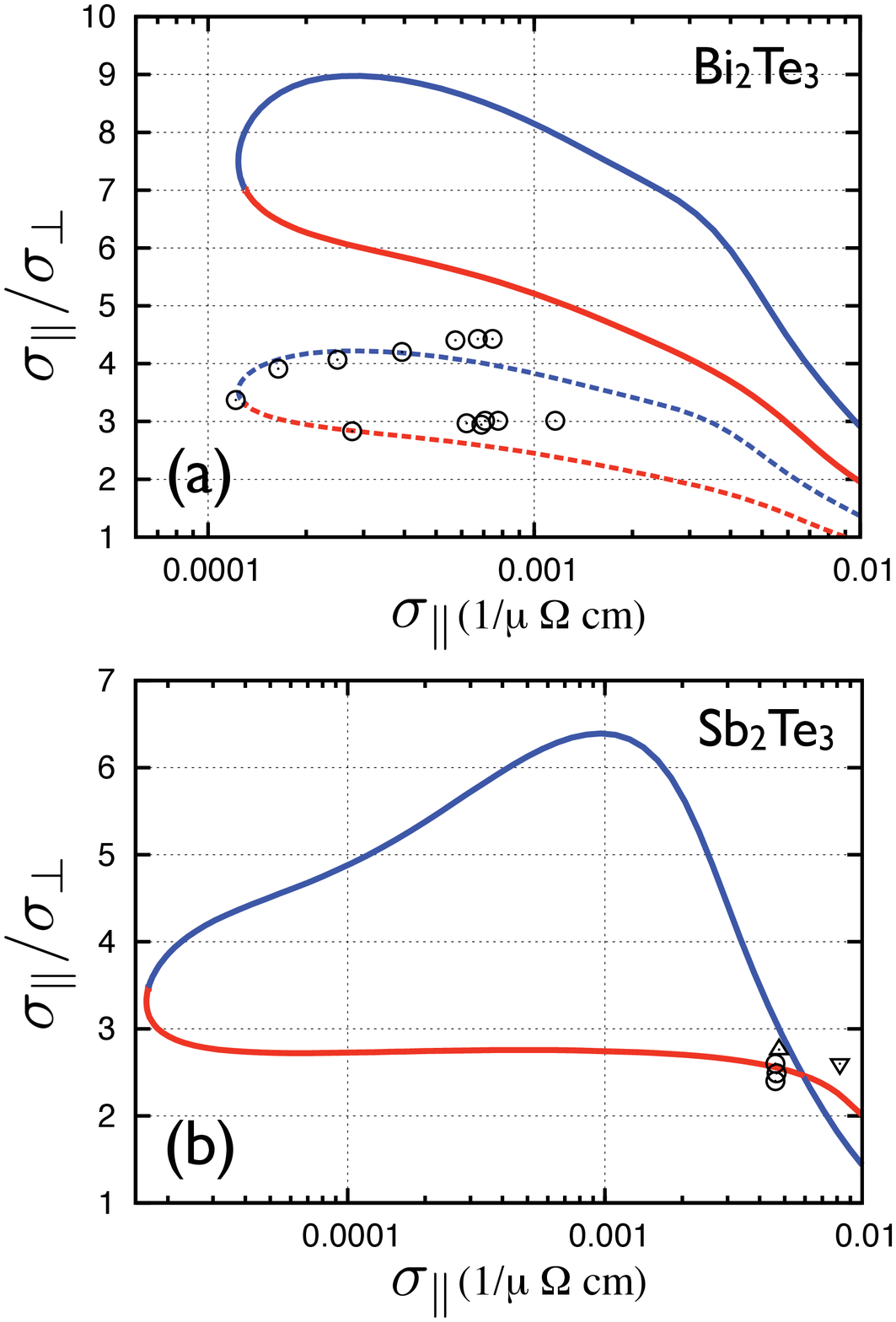}
\caption{\label{fig:2}(color online) Ratio \cip/\cpp of the electrical conductivites at 300K for unstrained bulk (a) \BiTe and (b) \SbTe. 
Electron doping refers to blue lines, while red lines refer to hole doping. 
The dashed lines in panel (a) present the ratio obtained with an anisotropic relaxation time $\nicefrac{\tau_{xx}}{\tau_{zz}}=0.47$, while all other results 
are obtained with an isotropic relaxation time. 
Experimental data (circles and triangles) from Ref.~\onlinecite{Delves:1961p7491,Jacquot:2010p15251,Langhammer:1982p15249} are given for comparison.}
\end{figure}
For this purpose the anisotropy of the electrical conductivity in dependence on the in-plane conductivity \cip for unstrained \BiTe and \SbTe is shown in \F{2}. 
The temperature is fixed at 300K, blue and red lines refer to electron- and hole-doping, respectively. 
From comparison with 
experimental data 
\footnote{The calculated dependencies of the electrical conductivity on the thermopower and 
the electrical conductivity on the applied doping were matched to fit experiments from Ref.~\onlinecite{Delves:1961p7491,Goldsmid:1958p15284,Langhammer:1982p15249}.}, 
the in-plane relaxation time is determined to be $\tau_{\|}=\unit[1.1\times 10 ^{-14}]{s}$ and $\tau_{\|}=\unit[1.2\times 10 ^{-14}]{s}$ for \BiTe and \SbTe, respectively. 
With that we find strong anisotropies for the electrical 
conductivity $\nicefrac{\sigma_{\|}}{\sigma_{\perp}} \gg 1$, clearly preferring the in-plane transport in both bulk tellurides. For 
the strongly suppressed cross-plane conduction p-type conduction is more favoured than n-type conduction. 
For \BiTe the pure band structure effects (solid lines in \F{2}(a)) overestimate the measured 
anisotropy ratio \cite{Delves:1961p7491} of the electrical conductivity. With an assumed anisotropy of the relaxation time of $\nicefrac{\tau_{\|}}{\tau_{\perp}}=0.47$ 
the experimental values are reproduced very well. That means, scattering effects strongly affect the transport and electrons travelling 
along the basal plane direction are scattered stronger than electrons travelling perpendicular to the basal plane. 
The origin of this assumed anisotropy has to be examined by defect calculations and resulting microscopic transition probabilities and state dependent 
mean free path vectors. It is well known, that in \BiTe mainly anti-site defects lead to the inherent 
conduction behaviour \cite{Mueller1998,MRowe:1995p15325,Cho:2011p15136}.
We have shown elsewhere \cite{PZ}, that the integration of the transport integrals eq.~\ref{Tcoeff} in anisotropic 
k-space requires large numeric effort. Tiny regions in k-space close to the band gap have to be scanned very carefully and the texture in k-space 
has a drastic influence on the obtained anisotropy values, if integrals are not converged with respect to the k-point density. 
As shown, some integration methods tend for the given k-space symmetry to underestimate the ratio $\nicefrac{\sigma_{\|}}{\sigma_{\perp}}$ 
in a systematic manner and therefore would shift anisotropy closer to the experimental observed values, without representing the real 
band structure effects.
For unstrained \BiTe the electrical conductivity anisotropy is highest for low values of \cip, i.e. small amounts of doping and bipolar conduction. 
For larger charge carrier concentrations, i.e. the chemical potential shifts deeper into either conduction or valence band, 
the in-plane conductivity \cip increases and the ratio $\nicefrac{\sigma_{\|}}{\sigma_{\perp}}$ decreases. Values for $\nicefrac{\sigma_{\|}}{\sigma_{\perp}}$ 
will lower from 7 to 2 for p-type conduction and 
9 to 3 for n-type conduction. However, cross-plane electrical transport is always more suppressed for n-type carrier conduction, which also holds for unstrained \SbTe. 
As shown in \F{2}(b) $\nicefrac{\sigma_{\|}}{\sigma_{\perp}}$ is almost doping independent for hole-doping, showing an anisotropy of around 2.7 in very 
good agreement with experiment (circle and triangles in \f{2} from Ref.~\onlinecite{Simon:1981p15250,Langhammer:1982p15249,Jacquot:2010p15251}). 
In this case no anisotropic relaxation times had to be assumed. For electron doping the ratio $\nicefrac{\sigma_{\|}}{\sigma_{\perp}}$ 
is clearly higher, evolving values of 3.5 to 6 for rising 
in-plane conductivity. The dependence of the anisotropy ratio on the applied doping, i.e. changing \cip, can be directly linked to the functional 
behaviour of the TDF near band edges, which is crucially influenced by the topology of the band structure \cite{BY}.
%

\section{\label{strained}Anisotropic thermoelectric properties of strained \BiTe and \SbTe}
Before the influence of in-plane strain on the resulting powerfactor will be discussed, we will first note on the strain induced 
changes of the components electrical conductivity and thermopower. In \F{3} the anisotropy of the electrical conductivity $\nicefrac{\sigma_{\|}}{\sigma_{\perp}}$ 
is shown for both \BiTe in the lattice constant of \SbTe, i.e. under biaxial compressive in-plane strain (\F{3}(a)), and 
\SbTe in the lattice constant of \BiTex, i.e. under biaxial tensile in-plane strain (\F{3}(b)). 
\begin{figure}[t]
\centering
\includegraphics[width=0.49\textwidth]{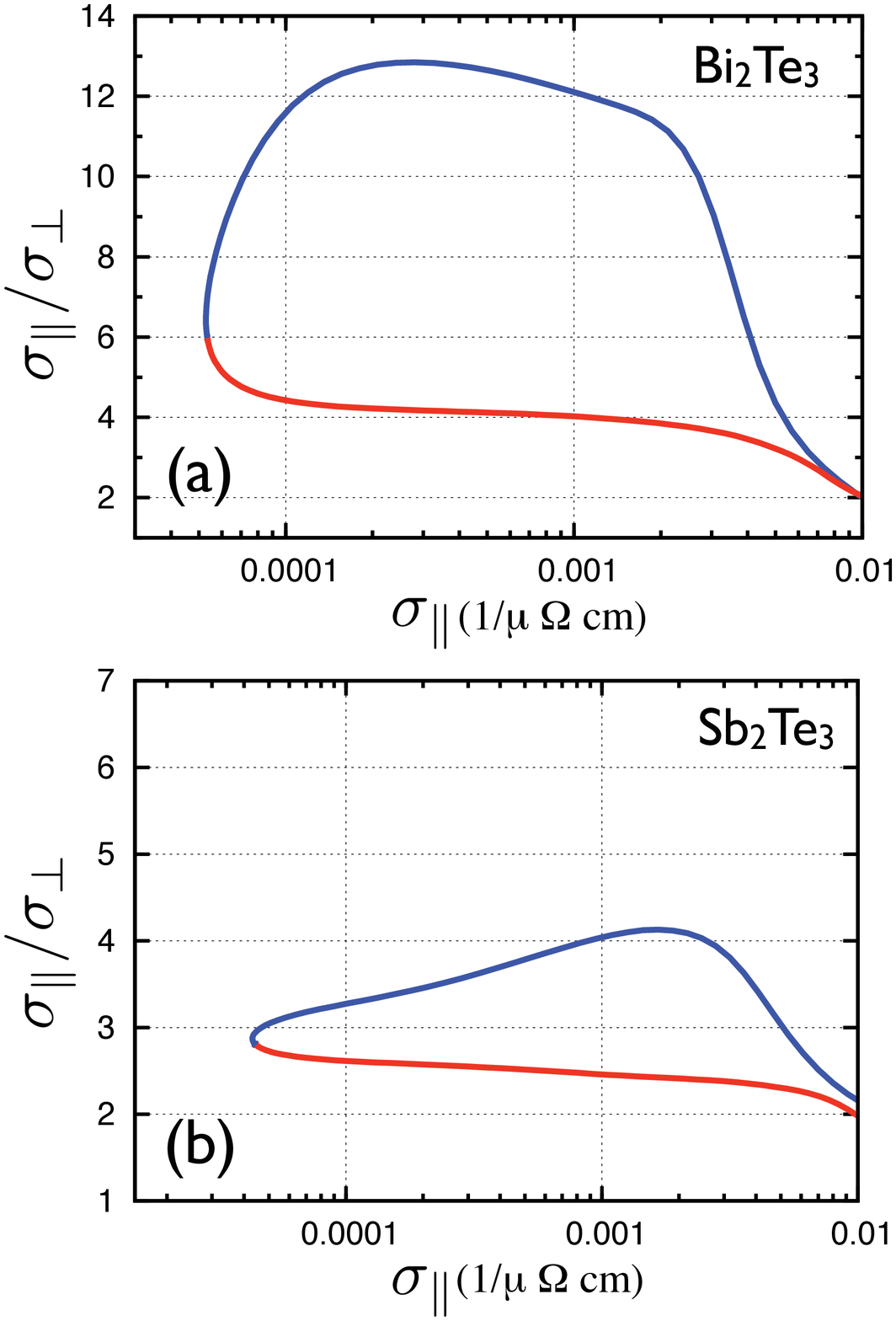}
\caption{\label{fig:3}(color online) Conductivity ratio \cip/\cpp of the electrical conductivites at 300K for bulk (a) \BiTe in the \SbTe structure and (b) \SbTe in the \BiTe structure. 
Electron doping refers to blue lines, while red lines refer to hole doping. Isotropic relaxation times of $\tau=\unit[1.1\times 10 ^{-14}]{s}$ 
and $\tau=\unit[1.2\times 10 ^{-14}]{s}$ for \cip and \cpp are assumed for \BiTe and \SbTe, respectively.}
\end{figure}

For \BiTe the compressive in-plane strain causes an increase of the the band gap by around  23\% yielding $E_g=\unit[129]{meV}$. 
While the anisotropy \ratio for hole doping (red lines in \f{3}(a)) decreases to around 4 and is almost constant 
under varying doping level, the ratio 
considerably raises under electron doping to values up to 13 for $\sigma_{_{\|}} \sim \unit[100 \dots 1000]{(\Omega \, cm){-1}}$, corresponding to 
electron charge carrier concentrations of $N = \unit[3 \dots 30 \times 10^{19}]{cm^{-3}}$. 
This concludes, that the cross-plane electrical conductivity 
of \BiTe under compressive in-plane strain will be noticeably enhanced for p-doping, but drastically suppressed for n-doping. 
Such a compressive in-plane strain 
could be introduced by either a substrate with smaller in-plane lattice constant, e.g. GaAs-[111] with $a=3.997${\AA }, or a considerable amount 
of \SbTe in the \SBSLx. 
For tensile in-plane strained \SbTe the impact on the electrical conductivity ratio \ratio is less prominent. As shown in \Ff	{3}(b) at hole doping
$\nicefrac{\sigma_{\|}}{\sigma_{\perp}} \sim 2.5$ is only marginally altered compared to the unstrained case (comp. \f{2}(b)). 
Meanwhile \ratio 
decreases noticeably for n-type doping yielding about 3 at low \cip and low electron charge 
carrier concentrations, and slightly higher values of $\nicefrac{\sigma_{\|}}{\sigma_{\perp}} \sim 4$ for higher doping. 
Overall, the tensile strain reduces 
the electrical conductivity anisotropy by a factor of about 1.5, directly leading to an enhanced electrical conductivity along the z-axis of single crystal \SbTe.
We note, that tensile in-plane strain opens the gap remarkably by around 56\% compared to the unstrained case to a value of $E_g=\unit[140]{meV}$. 
Furthermore, 
such tensile strain could be incorporated by using either a substrate with larger in-plane lattice constant, e.g. PbTe-[111] with $a=4.567${\AA }, or a higher fractional amount of \BiTe in \SBSLx. 
\begin{figure}[t]
\centering
\includegraphics[width=0.49\textwidth]{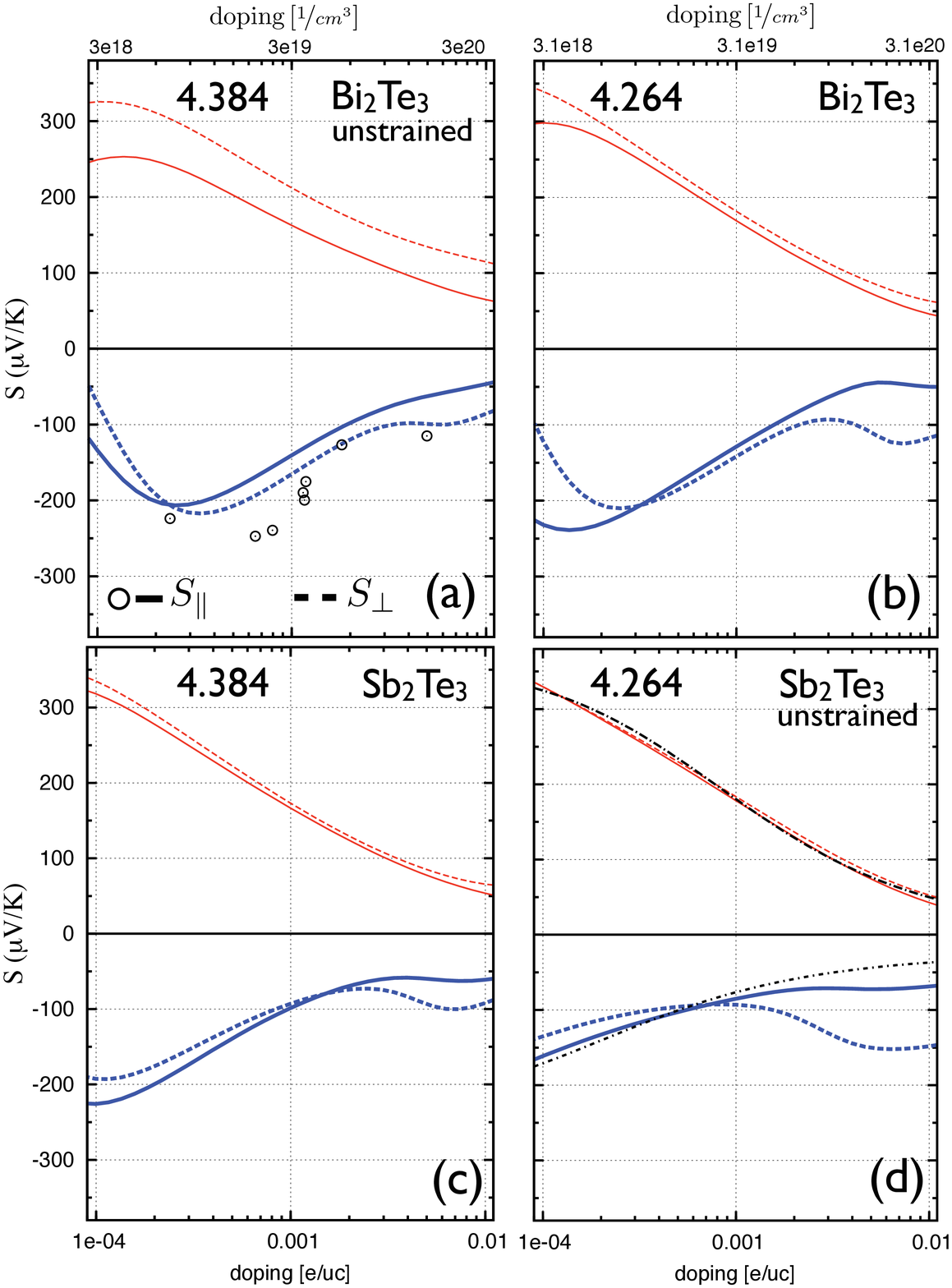}
\caption{\label{fig:6} (color online) In-plane (solid lines) and cross-plane (dashed lines) doping-dependent thermopower at 300K for (a) \BiTe in the \BiTe structure, (b) \BiTe in the \SbTe structure, (c) \SbTe in the \BiTe structure and (d) \SbTe in the \SbTe structure. Electron (hole) doping is presented as blue thick (red thin) line. 
The black (dashed-dotted) line in panel (d) shows the \textsc{Pisarenko}-dependence of the thermopower expected for parabolic bands. 
Experimental data (circles) from Ref.~\onlinecite{nurnus} is given for comparison. The charge carrier concentration 
is stated in units of $\unit[]{e/uc}$ ($\unit[]{1/cm^3}$) at the bottom (top) x-axis.}
\end{figure}
In \F{6}(a), (d) ((b), (c)) the doping dependent anisotropic thermopower of unstrained (strained) \BiTe and \SbTe at room temperature is shown, respectively. 
Blue thick (red thin) solid lines represent the in-plane thermopower \Sip under electron doping (hole doping). 
The corresponding cross-plane thermopower \Spp is shown as a dashed line. The black dashed-dotted lines in \f{6}(d) emphasize 
the expected doping dependent behaviour of the thermopower for parabolic bands, following the \textsc{Pisarenko}-relation\cite{Ioffe:1960}. 
For both tellurides we found, that the anisotropy of the thermopower shows a weak dependence on the strain state. 
However, for strained \BiTe (see \f{6}(b)) the thermopower anisotropy under hole doping almost vanishes, leading to $\Sipx \sim \Sppx$. 
It is worth noting, that the anisotropy of the thermopower is less pronounced for hole doping, than for electron doping for \BiTe and \SbTe in 
both strain states. As shown by the black dashed-dotted lines in \f{6}(d), the dependency 
of the thermopower on the charge carrier concentration differs from the \textsc{Pisarenko}-relation\cite{Ioffe:1960} under sufficient high electron doping. This 
indicates, that the nonparabolicity of the energy bands has a noticeable impact in the investigated doping regime and should not be 
omitted by applying parabolic band models.

Actually, changes for the absolute values of the thermopower can be found for both telluride systems under applied strain. 
\begin{figure}[t]
\centering
\includegraphics[width=0.49\textwidth]{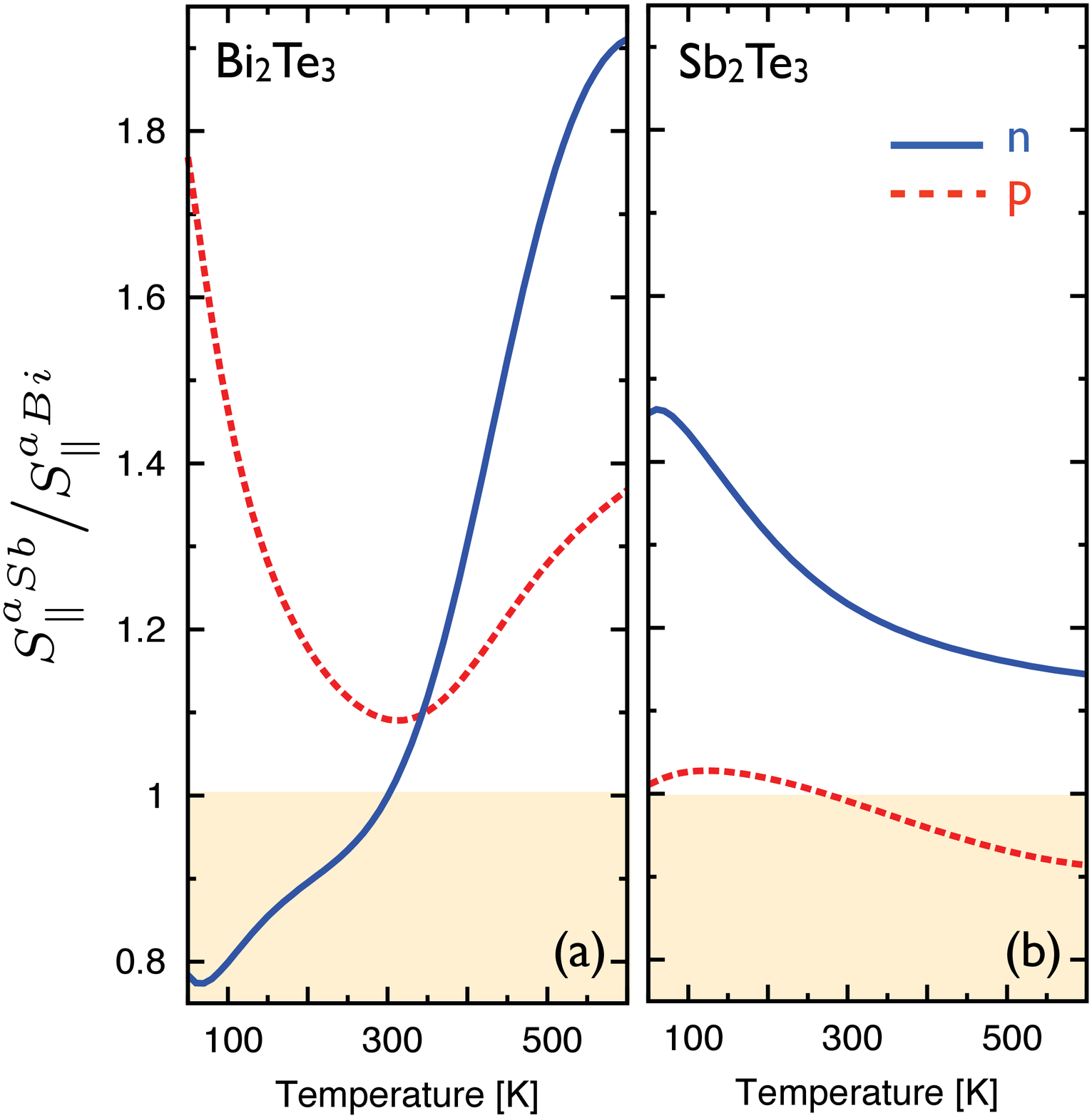}
\caption{\label{fig:4}(color online) Change of the in-plane thermopower \Sip under applied strain for (a) \BiTe and (b) \SbTe. Given is the ratio of 
\Sip in the "smaller" lattice of \SbTe divided by \Sip in the "larger" lattice of \BiTe. The doping was fixed to $N = \unit[1 \times 10^{19}]{cm^{-3}}$ for \BiTe 
and $N = \unit[1 \times 10^{20}]{cm^{-3}}$ for \SbTe. Solid blue (dashed red) lines refer to electron (hole) doping, respectively.}
\end{figure}
In \f{4} the relative change for the in-plane component \Sip for both tellurides under in-plane strain is given. 
To compare the changes with the lattice constant, we relate the in-plane thermopower \Sip at 
the smaller lattice constant $a_{SbTe}$ to the value at the larger lattice constant $a_{BiTe}$ for 
both compounds. The doping was fixed to 
$N = \unit[1 \times 10^{19}]{cm^{-3}}$ for \BiTe 
and $N = \unit[1 \times 10^{20}]{cm^{-3}}$ for \SbTe as done for \f{1}. 
\F{4}(a) shows, that in the relevant temperature range between 350K and 450K the thermopower 
increases for \BiTe under compressive strain for both p and n doping by about 15-20\%. 
For \SbTe a decrease is expected under tensile strain at electron doping and nearly no 
change under hole doping (see \F{4}(b)). 
With nearly all values above 1 for \BiTe, as well as for \SbTex, it is obvious, that higher values of the 
thermopower require a smaller unit cell volume. One can expect, that the volume decrease causes 
a larger density of states and thus a shift of the chemical potential towards the corresponding 
band edge, connected with an increase of the thermopower S. However Park \textit{et al.}\cite{Park:2010p11006} 
reported an unexpected increase of 16\% for the in-plane thermopower \Sip of \SbTe under p-doping 
(T=300K, $N = \unit[1.32 \times 10^{19}]{cm^{-3}}$) if the 
material is strained into the \BiTe structure. In the same doping and temperature regime we find a 
slight decrease of 4\% for \Sip. 

Comprising the statements on the electrical conductivity and the thermopower, the related powerfactor for both tellurides in their bulk lattice and in the 
strained state are compared in \f{5}. It is well known, that optimizing the powerfactor $\sigma S^2$ of a thermoelectric always involves a compromise 
on the electrical conductivity $\sigma$ and the thermopower $S$ \cite{Snyder:2008p240}. Due to the interdependence of $\sigma$ and $S$ it is not 
advisable to optimize the powerfactor by optimizing its parts. 
\begin{figure}[t]
\centering
\includegraphics[width=0.49\textwidth]{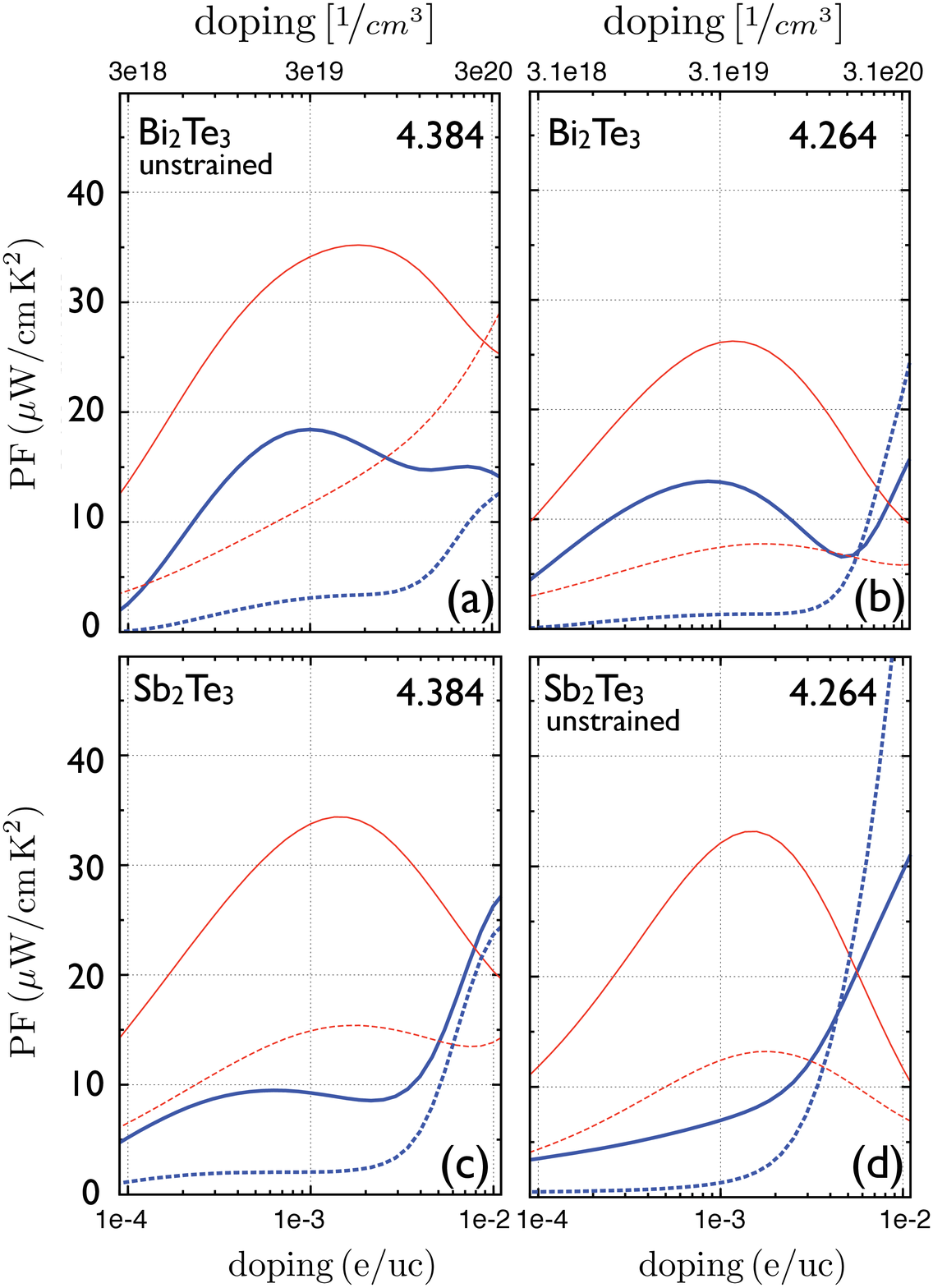}
\caption{\label{fig:5}(color online) In-plane (solid lines) and cross-plane (dashed lines) doping-dependent powerfactor at 300K for (a) \BiTe in the \BiTe structure, (b) \BiTe in the \SbTe structure, (c) \SbTe in the \BiTe structure and (d) \SbTe in the \SbTe structure. Electron (hole) doping is presented as blue thick (red thin) line. The charge carrier concentration 
is stated in units of $\unit[]{e/uc}$ ($\unit[]{1/cm^3}$) at the bottom (top) x-axis.}
\end{figure}
In \F{5}(a) and (d) the doping dependent anisotropic powerfactor of unstrained \BiTe and \SbTe at room temperature is shown, respectively. Blue thick (red thin) solid 
lines represent the in-plane powerfactor \PFip under electron doping (hole doping). The corresponding cross-plane powerfactor \PFpp is shown as a dashed line. 
Under p-doping both unstrained materials show a maximum powerfactor near carrier concentrations of
$N \sim \unit[4 \times 10^{19}]{cm^{-3}}$. Absolute values of $\unit[35]{\mu W/cm K^2}$ and $\unit[33]{\mu W/cm K^2}$ were found for 
unstrained \BiTe and \SbTex, respectively, which is in good agreement to experimental and theoretical findings \cite{nurnus,Bottner:2004p6591,Park:2010p11006}. 
Under electron doping the absolute values of \PFipx (thick blue lines in \f{5}) were found to be distinctly smaller. This is due to smaller absolute values of the thermopower for 
electron doping compared to hole doping (see \F{1}) and apparently smaller in-plane electrical conductivities \cip at fixed carrier concentrations. 
As a result, a powerfactor of $\unit[18]{\mu W/cm K^2}$ and $\unit[8]{\mu W/cm K^2}$ can be stated for unstrained \BiTe and \SbTex, respectively, under optimal 
electron doping. We notice, that the powerfactor for unstrained \SbTe is monotonically increasing for electron carrier concentrations of 
$N \sim \unit[6 \dots 30 \times 10^{19}]{cm^{-3}}$. This behaviour can be linked to a deviation of \Sip from the \textsc{Pisarenko}-relation under electron doping. 
While it is expected, that the thermopower will decrease for increasing carrier concentration, \Sip was found to be almost constant in an electron doping range of 
$N \sim \unit[6 \dots 30 \times 10^{19}]{cm^{-3}}$ (see \f{6}(d)). For the investigated electron doping range of 
$N \sim \unit[6 \dots 30 \times 10^{19}]{cm^{-3}}$ the chemical potential $\mu$ at 300K is located around $\unit[300 \dots 450]{meV}$ above the VBM. 
As can be seen from the band structure for unstrained \SbTe in \f{0}(b) (black, solid lines) flat non-parabolic bands near the high 
symmetry point Z dominate in this energy region and most likely lead to an increased thermopower. This feature is more pronounced for unstrained \SbTex, than for strained \SbTe (red, dashed lines in \f{0}(b)). Similar statements can be 
done for strained and unstrained \BiTe(see \f{0}(a)). 
We note, even though this picture is convincing, it is difficult to link such specific 
anomalies to the band structure on high symmetry lines, as the underlying TDF is an 
integral quantity over all occupied states in the BZ. 

Under applied in-plane compressive strain for \BiTe (ref. \F{5}(b)) and tensile strain for \SbTe (ref. \F{5}(c)) the obtained changes in the powerfactor 
are noticeable different for both tellurides. While for \BiTe a decrease of the maximal powerfactor \PFip of about 27\% and 23\% for 
n-doping and p-doping was found, the strain shows nearly no influence on the powerfactor for \SbTex. 
At a carrier concentration of about $N \sim \unit[3 \times 10^{19}]{cm^{-3}}$ the decrease in \PFip for \BiTe 
is about 17\% and 28\% for n- and p-doping, respectively, while in the work of Park \textit{et al.}\cite{Park:2010p11006} a slight increase of \PFip under strain and hole doping is reported. 
Obviously this tendency has to be understood by analyzing the constituent parts \cip and \Sip. 
For compressively strained \BiTe at a hole carrier concentration of about $N \sim \unit[3 \times 10^{19}]{cm^{-3}}$ the electrical conductivity 
decreases by about 39\% to $\unit[330]{(\Omega \, cm \, s)^{-1}}$. At the same time \Sip increases by about 9\%, as shown in \F{4}(a). 
This results in the overall decrease of about 28\% for \PFipx. Under electron doping of $N \sim \unit[3 \times 10^{19}]{cm^{-3}}$ no 
influence of strain could be found for \Sip at room temperature (see solid blue lines in \F{4}(a)). 
Thus, the decrease of \PFip under electron doping can be largely related 
to a decrease of the electrical conductivity under applied compressive strain. By detailed evaluation of the effective mass eigenvalues 
and eigenvectors we found a decrease of about 15\% for the in-plane electrical conductivity of \BiTe under applied strain in the low-temperature and 
low-doping limit\cite{PZ,BY}. The discussion can be made in the same manner for \SbTe \citep{PZ,BY}. 
The fact, that strain-induced effects in $\sigma$ and $S$ tend to compensate each other was already reported for the case of silicon \cite{Hinsche:2011p15276}. 

As mentioned before (summarized in \f{2} and \f{3}), we found a strong anisotropy in the electrical conductivity with $\nicefrac{\sigma_{\|}}{\sigma_{\perp}} \gg 1$. 
The clearly preferred in-plane transport in both bulk tellurides is also reflected in the cross-plane powerfactor \PFppx (dashed lines in \F{5}), which is clearly 
suppressed for all strain states. It is obvious that \PFpp is more suppressed for electron-, than for hole-doping.

Nonetheless, we want to include experimental findings for the thermal conductivity to our calculations, to give an estimation for the 
figure of merit $ZT$ in-plane and cross-plane. In Ref.~\onlinecite{Jacquot:2010p15251} $\kappa_{\|} = \unit[2.2]{W/m \,K}$, $\kappa_{\perp} = \unit[1.0]{W/m \,K}$, and 
$\kappa_{\|} = \unit[7.5]{W/m \,K}$, $\kappa_{\perp} = \unit[1.6]{W/m \,K}$ for unstrained \BiTe and \SbTe are given, respectively. 
With this we find maximal values for the figure of merit at room temperature and optimal hole doping of 
$ZT_{\|} \sim 0.48$ and $ZT_{\perp} \sim 0.41$ for unstrained \BiTe and $ZT_{\|} \sim 0.13$ and $ZT_{\perp} \sim 0.23$ for unstrained \SbTex. 
We note, that the figure of merit ZT maximizes at slightly lower carrier concentration than the powerfactor $\sigma S^2$ shown in \f{5}. 
This can be linked directly to an increasing electronic part of the 
thermal conductivity $\kappa_{el}$ with increasing carrier concentration \cite{Snyder:2008p240,Hinsche:2011p15276}.

\section{Conclusion} 
In the present paper the influence of in-plane strain on the thermoelectric transport properties of \BiTe and \SbTe is investigated. 
A focussed view on the influence of strain on the anisotropy of the electrical conductivity $\sigma$, thermopower $S$ and 
the related powerfactor $\sigma S^2$ could help to understand in-plane and cross-plane thermoelectric transport in nanostructured \BiTex/\SbTex-superlattices. 
Based on detailed \textit{ab initio} calculations we focussed mainly on band structure effects and their influence on the thermoelectric transport. 
For both tellurides no reasonable decrease of the anisotropy for $\sigma$ and $S$ could be found under strain, while in principle the 
anisotropy for $\sigma$ and $S$ is more pronounced under electron doping, than at hole doping. 
Thus a favoured thermoelectric transport along the z-direction of \BiTex/\SbTex-heterostructures due to 
superlattice-induced in-plane strain effects can be ruled out and a clear preference of p-type thermoelectric transport can be stated for \BiTe and \SbTe and their related epitaxial heterostructures. The absolute value of the in-plane thermopower \Sip was increased under reduced cell volume, which is in
contrast to recent findings by Park \textit{et al.}\cite{Park:2010p11006}. 

We found, that even if thermopower or electrical conductivity are enhanced or decreased via applied strain, they tend to compensate each other 
suppressing more distinct changes of the powerfactor under strain. 
We found the thermoelectrically optimal doping to be in the range of $N \sim \unit[3 \dots 6 \times 10^{19}]{cm^{-3}}$ for all considered systems. 
Our assumption of an anisotropic relaxation time for \BiTe states that already in the single crystalline system strong anisotropic scattering effects should play a role.

\begin{acknowledgments}
  This work was supported by the Deutsche
  Forschungsgemeinschaft, SPP 1386 `Nanostrukturierte Thermoelektrika: 
  Theorie, Modellsysteme und kontrollierte Synthese'. N. F. Hinsche is
  member of the International Max Planck Research School for Science
  and Technology of Nanostructures.

\end{acknowledgments}

\bibliography{draft_0.bbl}

\end{document}